\documentclass[pre,twocolumn,showpacs,showkeys,preprintnumbers,amsmath,amssymb]{revtex4}

\usepackage{graphicx}
\usepackage{dcolumn}
\usepackage{bm}
\usepackage{subfigure}
\usepackage{amssymb}
 \usepackage{graphics}
  \usepackage{epsfig}

\begin{document}


\title{Random field effects on the phase diagrams of spin-$1/2$ Ising model \\ on a honeycomb lattice}

\author{Yusuf Y\"{u}ksel}
\author{\"{U}mit Ak{\i}nc{\i}}
\author{Hamza Polat}\email{hamza.polat@deu.edu.tr}

\affiliation{ Department of Physics, Dokuz Eyl\"{u}l University,
TR-35160 Izmir, Turkey} \affiliation{ Dokuz Eyl\"{u}l University,
Graduate School of Natural and Applied Sciences, Turkey}
\date{\today}

\begin{abstract}
Ising model with quenched random magnetic fields is examined for single Gaussian, bimodal and double Gaussian random field distributions by introducing an effective field approximation that takes into account the correlations between different spins that emerge when expanding the identities. Random field distribution shape dependence of the phase diagrams, magnetization and internal energy is investigated for a honeycomb lattice with a coordination number $q=3$. The conditions for the occurrence of reentrant behavior and tricritical points on the system are also discussed in detail.
\end{abstract}



\maketitle

\section{Introduction}
Ising model \cite{ising,onsager} which was originally introduced as a model describing the phase transition properties of ferromagnetic materials has been widely examined in statistical mechanics and condensed matter physics. In the course of time, basic prescience of this simple model has been improved by introducing new concepts such as disorder effects on the critical behavior of the systems in question. Ising model in a quenched random field (RFIM) which has been studied over three decades is an example of this situation. The model which is actually based on the local fields acting on the lattice sites which are taken to be random according to a given probability distribution was introduced for the first time by Imry and Ma \cite{imry_ma}. Subsequently, a great many of theoretical and experimental works have paid attention on the RFIM and quite noteworthy results have been obtained. For instance, it has been shown that diluted antiferromagnets such as $Fe_{x}Zn_{1-x}F_{2}$\cite{belanger,king}, $Rb_{2}Co_{x}Mg_{1-x}F_{4}$\cite{ferreira,yoshizawa} and $Co_{x}Zn_{1-x}F_{2}$\cite{yoshizawa}  in a uniform magnetic field just correspond to a ferromagnet in a random uniaxial magnetic field \cite{fishman,cardy}.

Following studies have been devoted to investigate the phase diagrams of these systems in depth and in the mean field level it was found that different random field distributions lead to different phase diagrams. For example, using a Gaussian probability distribution Schneider and Pytte\cite{schneider_pytte} have shown that phase diagrams of the model exhibit only second order phase transition properties. On the other hand, Aharony \cite{aharony} and Matthis \cite{matthis} have introduced bimodal and trimodal distributions, respectively and they have reported the observation of tricritical behavior. In order to clarify this controversial situation, the problem have been investigated by a variety of theoretical works such as effective field theory (EFT) \cite{albuquerque1,albuquerque2,bobak,borges,jascur,liang1,liang2,qiu,sarmento,sebastianes,kaneyoshi}, Monte Carlo (MC) simulations \cite{gawlinski,landau,machta,mackenzie,grest,weizenmann}, mean field (MF) approximation \cite{galam,kaufmann,milman,salinas,hadjiagapiou,yokota}, pair approximation (PA) \cite{albayrak,canko}, Bethe-Peierls approximation (BPA) \cite{wohlman} and series expandion (SE) \cite{gofman} method. Moreover, recently phase transition properties of RFIM with symmetric double \cite{crokidakis} and triple \cite{salmon} Gaussian random fields have also been studied by means of a replica method and a rich variety of phase diagrams have been presented.

EFT has been widely used in the literature and some interesting results have been reported. For example, de Albuquerque et al. \cite{albuquerque1} have studied the behavior of the spin-1/2 RFIM on a honeycomb lattice with a bimodal field distribution and they have observed the absence of a tricritical point. Liang and co-workers \cite{liang2} examined spin-3/2 system with a random field on honeycomb, square, and simple-cubic lattices, respectively and reported the existance of tricritical point and the reentrant phenomena for a system with any coordination number $q$, when the applied random field is bimodal. Sarmento and Kaneyoshi \cite{sarmento} have studied the phase diagrams of a transverse spin-1/2 Ising model with a random field and found a reentrant behavior of second order. They have not observed any tricritical point for the system with a coordination number $q=3$ when the applied field is bimodal. Sebastianes and Figueiredo \cite{sebastianes} have studied a semi-infinite simple cubic lattice with a trimodal random field distribution and they have observed first order transitions on the system.

Conventional EFT approximations include spin-spin correlations resulting from the usage of the Van der Waerden identities and provide results that are superior to those obtained within the traditional MFT. However, these conventional EFT approximations are not sufficient enough to improve the results, due to the usage of a decoupling approximation (DA) that neglects the correlations between different spins that emerge when expanding the identities. Therefore, taking these correlations into consideration will improve the results of conventional EFT approximations. In order to overcome this point, recently we proposed an approximation that takes into account the correlations between different spins in the cluster of considered lattice \cite{polat,canpolat,yuksel_1,yuksel_2,yuksel_3}. Namely, an advantage of the approximation method proposed by these studies is that no uncontrolled decoupling procedure is used for the higher order correlation functions. On the other hand, as far as we know EFT studies in the literature dealing with RFIM are based only on  discrete probability distributions (bimodal or trimodal). Hence, in this work we intent to study the phase diagrams of the RFIM with single Gaussian, bimodal and double Gaussian random field distributions.

Organization of the paper is as follows: In section \ref{formulation} we briefly present the formulations. The results and discussions are presented in section \ref{results}, and finally section \ref{conclude} contains our conclusions.

\section{Formulation}\label{formulation}
We consider a two-dimensional lattice which has $N$ identical spins arranged. We define a
cluster on the lattice which consists of a central spin labeled $S_{0}$, and $q$ perimeter spins being the
nearest neighbors of the central spin. The cluster consists of $(q+1)$ spins being independent
from the value of $S$. The nearest neighbor spins are in an effective field produced by the outer
spins, which can be determined by the condition that the thermal average of the central spin is
equal to that of its nearest neighbor spins. The Hamiltonian describing our model is
\begin{equation}\label{eq1}
H=-J\sum_{<i,j>}S_{i}^{z}S_{j}^{z}-\sum_{i}h_{i}S_{i}^{z},
\end{equation}
where the first term is a summation over the nearest neighbor spins with $S_{i}^{z}=\pm1$ and
the second term represents the Zeeman interactions on the lattice. Random magnetic fields are distributed according to a given Gaussian distribution function.
Present study deals with three kinds of Gaussian field distribution. Namely, a normal distribution which is defined as
\begin{equation}\label{eq2b}
P(h_{i})=\left(\frac{1}{2\pi\sigma^{2}}\right)^{1/2}\exp\left[-\frac{h_{i}^{2}}{2\sigma^{2}}\right],
\end{equation}
with zero mean and width $\sigma$, a bimodal discrete distribution
\begin{equation}\label{eq2a}
P(h_{i})=\frac{1}{2}\left[\delta(h_{i}-h_{0})+\delta(h_{i}+h_{0})\right],
\end{equation}
where half of the lattice sites subject to a magnetic field $h_{0}$ and the remaining lattice sites have a field $-h_{0}$,  and a double peaked Gaussian distribution
\begin{eqnarray}\label{eq2c}
\nonumber P(h_{i})&=&\frac{1}{2}\left(\frac{1}{2\pi\sigma^{2}}\right)^{1/2}\left\{\exp\left[-\frac{(h_{i}-h_{0})^{2}}{2\sigma^{2}}\right]\right.\\
&&\left.+ \exp\left[-\frac{(h_{i}+h_{0})^{2}}{2\sigma^{2}}\right]\right\}
\end{eqnarray}
In a double peaked distribution defined in equation (\ref{eq2c}), random fields $\pm h_{0}$ are distributed with equal probability and the form of the distribution depends on the $h_{0}$ and $\sigma$ parameters, where $\sigma$ is the width of the distribution.

According to Callen identity \cite{callen} for the spin-1/2 Ising ferromagnetic system with the coordination number $q$, the thermal average of
the spin variables at the site $i$ is given by
\begin{equation}\label{eq3}
\left\langle \{f_{i}\}S_{i}^{z}\right\rangle=\left\langle \{f_{i}\}\tanh\left[\beta\left(\sum_{j} J_{ij}S_{j}+h_{i}\right)\right]\right\rangle,
\end{equation}
where $\nabla=\partial/\partial x$ is a differential operator, $j$ expresses the nearest neighbor sites of the central spin and $\{f_{i}\}$ can be any function of the Ising variables as long as it is not a function of the site. From equation (\ref{eq3}) with $f_{i}=1$, the thermal and random-configurational averages of a central spin can be represented in the form for honeycomb $(q=3)$ lattice by introducing the differential operator technique \cite{honmura_kaneyoshi,kaneyoshi_1}
\begin{eqnarray}\label{eq4}
\nonumber \left\langle\left\langle S_{0}^{z}\right\rangle\right\rangle_{r}&=&\left\langle\left\langle \prod_{j=1}^{q}\left[\cosh(J\nabla)+S_{j}^{z}\sinh(J\nabla)\right]\right\rangle\right\rangle_{r}\\
&&\ \ \ \ \ \ \ \ \ \ \ \ \ \ \ \ \ \ \ \ \ \ \ \ \ \ \ \ \ \ \ \ \ \ \ \ \times F(x)|_{x=0},
\end{eqnarray}
where the inner $\langle...\rangle$ and the outer $\langle...\rangle_{r}$ brackets represent the thermal and configuratio-\newline nal averages, respectively. The function $F(x)$ in equation (\ref{eq4}) is defined by
\begin{equation}\label{eq5}
F(x)=\int dh_{i}P(h_{i})\tanh[\beta(x+h_{i})],
\end{equation}
and it has been calculated by numerical integration and by using the distribution functions defined in equations (\ref{eq2b}), (\ref{eq2a}) and (\ref{eq2c}).
By expanding the right hand side of equation (\ref{eq4}) we get the longitudinal spin correlation as
\begin{eqnarray}\label{eq6}
\nonumber \langle\langle S_{0}^{z}\rangle\rangle_{r}&=&k_{0}+3k_{1}\langle\langle S_{1}\rangle\rangle_{r}+3k_{2}\langle\langle S_{1}S_{2}\rangle\rangle_{r}\\
&&+k_{3}\langle\langle S_{1}S_{2}S_{3}\rangle\rangle_{r}.
\end{eqnarray}
The coefficients in equation (\ref{eq6}) are defined as follows
\begin{eqnarray}\label{eq7}
\nonumber
k_{0}&=&\cosh(J\nabla)F(x)|_{x=0},\\
\nonumber
k_{1}&=&\cosh^{2}(J\nabla)\sinh(J\nabla)F(x)|_{x=0},\\
\nonumber
k_{2}&=&\cosh(J\nabla)\sinh^{2}(J\nabla)F(x)|_{x=0},\\
k_{3}&=&\sinh^{3}(J\nabla)F(x)|_{x=0}.
\end{eqnarray}
Next, the average value of the perimeter spin in the system can be written as follows and it is found as
\begin{eqnarray}\label{eq8}
\nonumber
m_{1}=\langle\langle S_{1}^{z}\rangle\rangle_{r} &=&\langle\langle \cosh(J\nabla)+S_{0}^{z}\sinh(J\nabla)\rangle\rangle_{r} F(x+\gamma),\\
&=&a_{1}+a_{2}\langle\langle S_{0}\rangle\rangle_{r}.
\end{eqnarray}
For the sake of simplicity, the superscript $z$ is omitted from the right hand sides of equations (\ref{eq6}) and (\ref{eq8}).
The coefficients in equation (\ref{eq8}) are defined as
\begin{eqnarray}\label{eq9}
\nonumber
a_{1}&=&\cosh(J\nabla)F(x+\gamma)|_{x=0},\\
a_{2}&=&\sinh(J\nabla)F(x+\gamma)|_{x=0}.
\end{eqnarray}
In equation (\ref{eq9}), $\gamma=(q -1)A$ is the effective field produced by the $(q-1)$ spins outside of the system and $A$ is an unknown parameter to be determined self-consistently. Equations (\ref{eq6}) and (\ref{eq8}) are the fundamental correlation functions of the system. When the right-hand side of equation (\ref{eq4}) is expanded, the multispin correlation functions appear. The simplest approximation, and one of the most frequently adopted is to decouple these correlations according to
\begin{equation}\label{eq10}
\left\langle\left\langle
S_{i}^{z}(S_{j}^{z})^{2}...S_l^z\right\rangle\right\rangle_{r}\cong\left\langle\left\langle
S_i^z\right\rangle\right\rangle_{r}\left\langle\left\langle
(S_j^z)^{2}\right\rangle\right\rangle_{r}...\left\langle\left\langle
S_l^z\right\rangle\right\rangle_{r},
\end{equation}
for $i\neq j \neq...\neq l$ \cite{tamura_kaneyoshi}. The main difference of the method used in this study from the other approximations in the literature emerges in comparison with any decoupling approximation (DA) when expanding the right-hand side of
equation (\ref{eq4}). In other words, one advantage of the approximation method used in this study is that no uncontrolled decoupling procedure is used for the higher order correlation functions.

For spin-1/2 Ising system in a random field with $q =3$, taking equations (\ref{eq6}) and (\ref{eq8}) as a basis we derive a set of linear equations of the spin correlation functions which interact in the system. At this point, we assume that $(i)$ the correlations depend only on the distance between the spins and $(ii)$ the average values of a central spin and its nearest-neighbor spin (it is labeled as the perimeter spin) are equal to each other with the fact that, in the matrix representations of spin operator $\hat{S}$, the spin-1/2 system has the property $(S_{\delta}^{z})^{2}=1$ . Thus, the number of linear equations obtained for the system with $q=3$ reduces to six and the complete set is as follows
\begin{eqnarray}\label{eq11}
\nonumber
\langle\langle S_{0}\rangle\rangle_{r}&=&k_{0}+3k_{1}\langle\langle S_{1}\rangle\rangle_{r}+3k_{2}\langle\langle S_{1}S_{2}\rangle\rangle_{r}\\
\nonumber && +k_{3}\langle\langle S_{1}S_{2}S_{3}\rangle\rangle_{r}\\
\nonumber
\left\langle\left\langle S_{0}S_{1}\right\rangle\right\rangle_{r}&=&3k_{1}+(k_{0}+3k_{2})\left\langle\left\langle S_{1} \right\rangle\right\rangle_{r}+k_{3}\left\langle\left\langle S_{1}S_{2}\right\rangle\right\rangle_{r}\\
\nonumber
\left\langle\left\langle S_{0}S_{1}S_{2}\right\rangle\right\rangle_{r}&=&(3k_{1}+k_{3})\left\langle\left\langle S_{1}\right\rangle\right\rangle_{r}+(k_{0}+3k_{2})\left\langle\left\langle S_{1}S_{2}\right\rangle\right\rangle_{r}\\
\nonumber
\left\langle\left\langle S_{1}\right\rangle\right\rangle_{r}&=&a_{1}+a_{2}\left\langle\left\langle S_{0}\right\rangle\right\rangle_{r}\\
\nonumber
\left\langle\left\langle S_{1}S_{2}\right\rangle\right\rangle_{r}&=&a_{1}\left\langle\left\langle S_{1}\right\rangle\right\rangle_{r}+a_{2}\left\langle\left\langle S_{0}S_{1}\right\rangle\right\rangle_{r}\\
\left\langle\left\langle S_{1}S_{2}S_{3}\right\rangle\right\rangle_{r}&=&a_{1}\left\langle\left\langle S_{1}S_{2}\right\rangle\right\rangle_{r}+a_{2}\left\langle\left\langle S_{0}S_{1}S_{2}\right\rangle\right\rangle_{r}.
\end{eqnarray}
If equation (\ref{eq11}) is written in the form of a $6\times6$ matrix and solved in terms of the variables $x_{i}[(i=1,2,...,6)(e.g., x_{1}=\langle\langle S_{0}^{z}\rangle\rangle_{r}, x_{2}=\langle\langle S_{0}S_{1}\rangle\rangle_{r},..., x_{6}=\langle\langle S_{1}S_{2}S_{3}\rangle\rangle_{r})]$ of the linear equations, all of the spin correlation functions can be easily determined as functions of the temperature and random field parameters $\sigma$ and $h_{0}$. Since the thermal and configurational averages of
the central spin is equal to that of its nearest-neighbor spins within the present method then the unknown parameter $A$ can be numerically determined by the relation
\begin{equation}\label{eq12}
\langle\langle S_{0}\rangle\rangle_{r}=\langle\langle
S_{1}\rangle\rangle_{r} \qquad {\rm{ or }}\qquad x_{1}=x_{4}.
\end{equation}
By solving equation (\ref{eq12}) numerically at a given fixed set of Hamiltonian/random field parameters we obtain the parameter $A$. Then we use the numerical values of $A$ to obtain the spin correlation functions which can be found from equation (\ref{eq11}). Note that $A=0$ is always the root of equation (\ref{eq12}) corresponding to the disordered state of the system. The nonzero root of $A$ in equation (\ref{eq12}) corresponds to the long range ordered state of the system. Once the spin correlation functions have been evaluated then we can give the numerical results for the thermal and magnetic properties of the system. Since the effective field $\gamma$ is very small in the vicinity of $k_{B}T_{c}/J$, we can obtain the critical temperature for the fixed set of Hamiltonian parameters by solving equation (\ref{eq12}) in the limit of $\gamma\rightarrow0$ then we can construct the whole phase diagrams of the system. Depending on the Hamiltonian parameters, there may be two solutions (i.e. two critical temperature values satisfy the equation (\ref{eq12})) corresponding to the first/second and second order phase transition points, respectively. We determine the type of the transition by looking at the temperature dependence of magnetization for selected values of Hamiltonian parameters.

\section{Results and Discussion}\label{results}

In this section, we discuss how the type of random field distribution effects the phase diagrams of the system. Also, in order to clarify the type of the transitions in the system, we give the temperature dependence of the order parameter.

\subsection{Phase diagrams of single Gaussian distribution}
The form of single Gaussian distribution which is defined in equation (\ref{eq2b}) is governed by only one parameter $\sigma/J$ which is the width of the distribution.
On the left panel in Fig.1a, we show the phase diagram of spin-1/2 system on a honeycomb lattice in $(k_{B}T_{c}/J-\sigma/J)$ plane. We can clearly see that as $\sigma/J$ increases then the width of the
\begin{figure}[!h]\label{fig1}
\center\subfigure[\hspace{0 cm}] {\includegraphics[width=8cm]{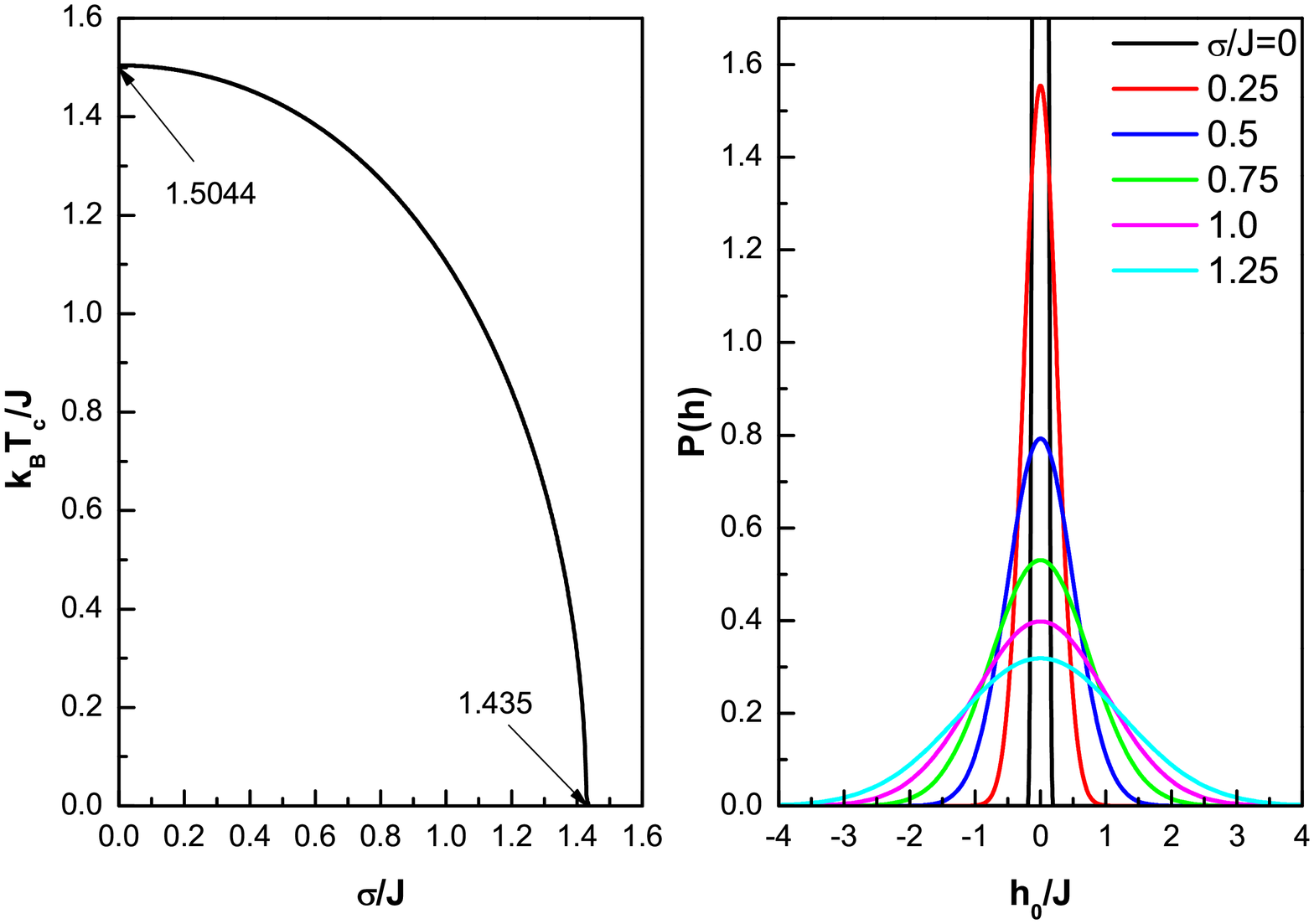}}\\
\subfigure[\hspace{0 cm}] {\includegraphics[width=7cm]{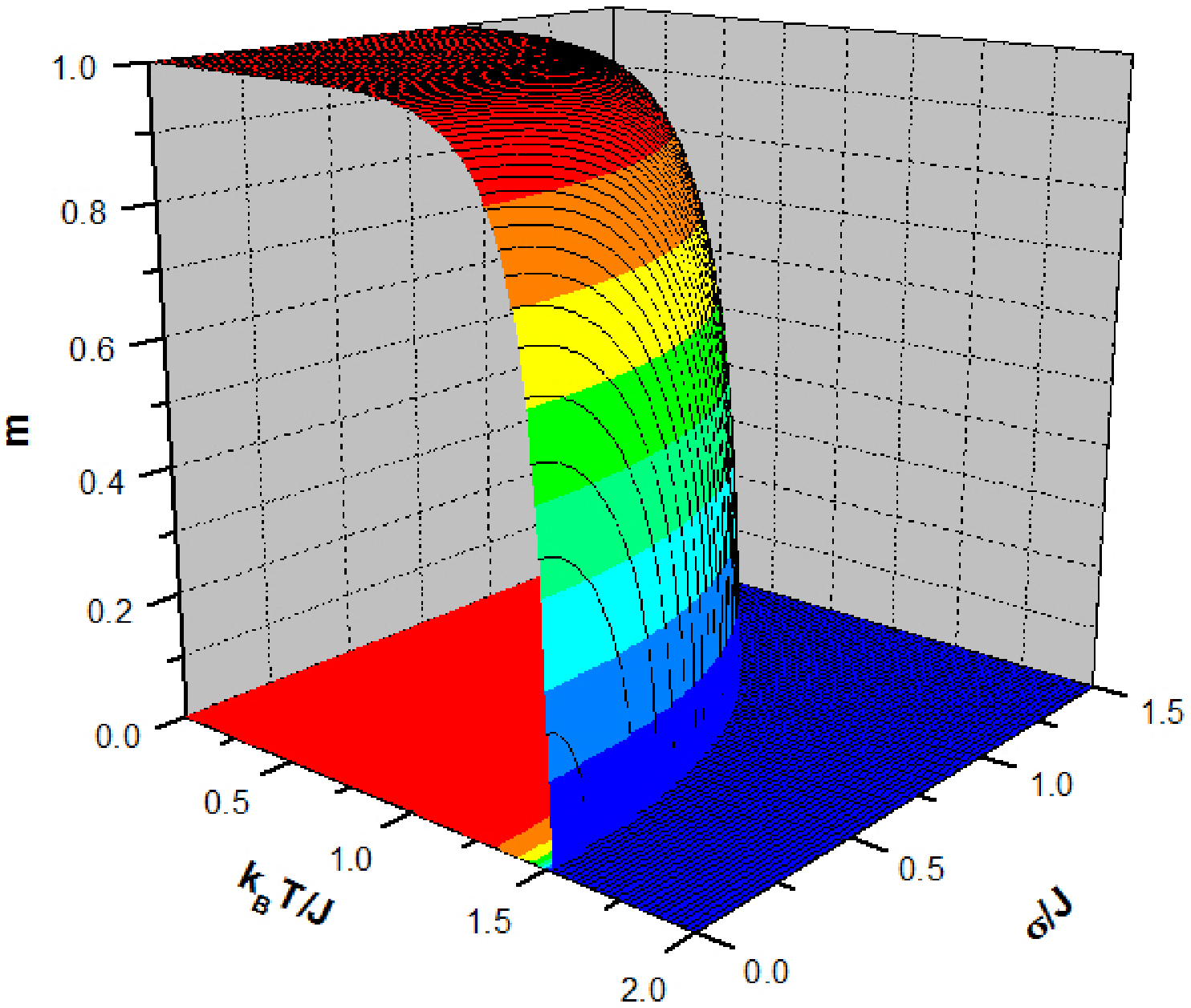}}
\subfigure[\hspace{0 cm}] {\includegraphics[width=7cm]{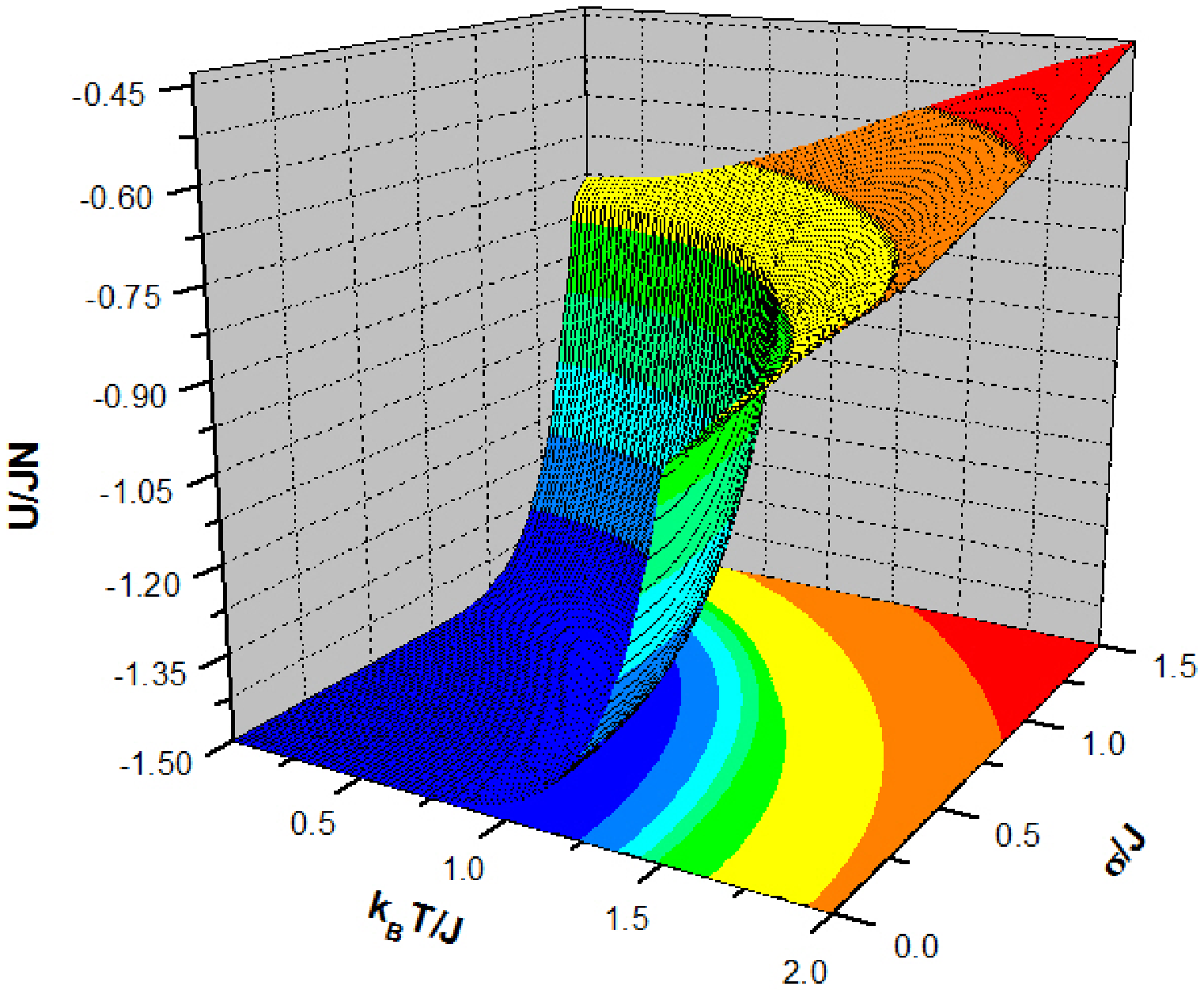}}
\caption{(a) Phase diagram (left panel) of system corresponding to single Gaussian distribution in $(k_{B}T_{c}/J-\sigma/J)$ plane and variation of single Gaussian distribution with $h_{0}/J$ for selected values of $\sigma/J$ (right panel). (b) Magnetization and (c) internal energy surfaces on $(m-k_{B}T/J-\sigma/J)$ and $(U/JN-k_{B}T/J-\sigma/J)$  planes, respectively.}
\end{figure}
distribution function gets wider and randomness effect of magnetic field distribution on the system becomes significantly important (see the right panel in Fig. 1a). Therefore, increasing $\sigma/J$ value causes a decline in the critical temperature $k_{B}T_{c}/J$ of the system. $k_{B}T_{c}/J$ value in the absence of any randomness i.e. when $\sigma/J=0$ is obtained as $k_{B}T_{c}/J=1.5044$ and this value just corresponds to the critical temperature of the pure system. This value is in a good agreement with the exact result $1.519$ \cite{fisher}. We also note that the critical temperature of the system reaches to zero at $\sigma/J=1.435$. Besides, we have not observed any reentrant/tricritical behavior for single Gaussian distribution. The system undergoes only second order phase transition as depicted in Figs. 1b and 1c. In Figs. 1b and 1c, we show the order parameter (magnetization) and the internal energy surfaces in $(m-k_{B}T/J-\sigma/J)$ and $(U/JN-k_{B}T/J-\sigma/J)$ planes, respectively. As we can clearly see from Fig. 1b, as the temperature increases then the magnetization of the system tends to decrease from its saturation value and falls to zero at a second order phase transition temperature. Furthermore, in the presence of any random field corresponding to equation (\ref{eq2b}), increasing $\sigma/J$ values give rise to a decrement in the critical temperature of the system and hence the ferromagnetic region gets narrower. The internal energy as a function of the temperature and distribution width $\sigma/J$ which is depicted in Fig. 1c also agrees with the observations mentioned above. Thus, we see that our results support the previously published works in the literature.

\subsection{Phase diagrams of bimodal distribution}
Next, in order to investigate the effect of the bimodal random fields defined in equation (\ref{eq2a}) on the phase diagrams of the system, we show the phase diagram on the $(k_{B}T_{c}/J-h_{0}/J)$ plane and temperature dependence of magnetization $m$ on the left and right hand side panels in Fig. 2, respectively. The right panel in Fig. 2 shows three different magnetization profiles. Namely, a second order transition occurs between
\begin{figure}[!h]\label{fig2}
\center
\includegraphics[width=8cm]{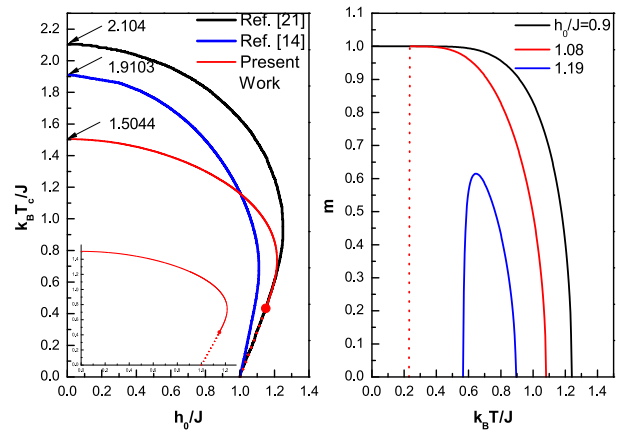}\\
\caption{Phase diagram in $(k_{B}T_{c}-h_{0}/J)$ plane (left panel) corresponding to bimodal field distribution obtained from present work and results of Ref.\cite{albuquerque2} and Ref.\cite{sarmento} and temperature dependence of magnetization (right panel) with some selected values of $h_{0}/J$. Solid and dotted lines correspond to second and first order phase transitions and solid circle denotes the tricritical points.}\label{fig2}
\end{figure}
paramagnetic and ferromagnetic phases for $h_{0}/J=0.9$, a first order phase transition (dotted line on the right panel in Fig.2) is followed by a second order phase transition for $h_{0}/J=1.08$ and a reentrant behavior of second order appears for $h_{0}/J=1.19$. On the left panel in Fig. 2, we also compare our results by the other works based on EFT in the literature. In Refs.\cite{albuquerque2} and \cite{sarmento} the authors reported a reentrant behavior for spin-1/2 system on a honeycomb lattice $(q=3)$ with bimodal random fields but they did not observe a tricritical point. On the other hand, as seen from Fig. 2 the system exhibits a tricritical behavior as well as reentrant phenomena within the present work. According to our calculations, reentrant phenomena and the first order phase transitions can be observed in the range of $1.0<h_{0}/J<1.215$ and $1.0<h_{0}/J<1.151$, respectively. Furthermore, for $h_{0}/J=0$ (i.e. pure system again) the critical temperature $k_{B}T_{c}/J=1.5044$ can be compared with $1.9103$ of Ref.\cite{albuquerque2} and $2.104$ of Ref.\cite{sarmento} and the exact result $1.519$ of Ref.\cite{fisher}. Hence, it is obvious that the present method is superior to the other EFT methods in the literature. The reason is due to the fact that, in contrast to the previously published works mentioned above there is no uncontrolled decoupling procedure used for the higher order correlation functions within the present approximation.

\subsection{Phase diagrams of double Gaussian distribution}
Double Gaussian distribution in equation (\ref{eq2c}) have not yet been examined within the framework of EFT in the literature. Therefore, it would be interesting to investigate the phase diagrams of the system with random fields corresponding to equation (\ref{eq2c}). Now
the shape of the random fields is governed by two parameters $h_{0}/J$ and $\sigma/J$. In Fig. 3a, right panel shows three different magnetization profiles for $\sigma/J=0.1$. Namely, a second order transition occurs between paramagnetic and ferromagnetic phases for $h_{0}/J=1.0$, a first order phase transition (dotted line on the right panel in Fig. 3a) is followed by a second order phase transition for $h_{0}/J=1.11$ and a reentrant behavior of second order appears for $h_{0}/J=1.2$. The left panel in Fig. 3a depicts the phase diagram of the system in $(k_{B}T_{c}/J-h_{0}/J)$ plane with selected values of $\sigma/J$ where dotted lines correspond to first order transitions and solid circles represent the tricritical points. As seen from the figure, the system exhibits tricritical points and reentrant phenomena for weak randomness and as $\sigma/J$ increases then
reentrant phenomena and tricritical behavior disappear. Furthermore, the critical temperature of the system decreases and ferromagnetic phase region gets narrower with increasing trend in $\sigma/J$. On the left panel in Fig. 3b, we investigate the phase diagram of the system in $(k_{B}T_{c}/J-\sigma/J)$ plane with selected values of $h_{0}/J$. As seen from the figure, for the values of $h_{0}/J=0, 0.5, 0.7$ and $1.0$ the system undergoes a second order phase transition between paramagnetic and ferromagnetic phases at a critical temperature which decreases with increasing values of $h_{0}/J$. For the values of $h_{0}/J=1.21, 1.22, 1.24$ and $1.26$ the system exhibits a reentrant behavior of second order and the transition lines exhibit a bulge which gets smaller with increasing values of $h_{0}/J$ which means that ferromagnetic phase region gets narrower. Besides, the magnetization profiles shown on the right panel in Fig. 3b for $h_{0}/J=1.21$ with $\sigma/J=0.2, 0.4$ and $0.6$ show that reentrant behavior disappears with increasing trend in $\sigma/J$. Random field distribution shapes  corresponding to the magnetization profiles in Fig. 3a and Fig. 3b are also illustrated in Fig. 3c.

Finally, in Fig. 4 we show the phase diagram in $(h_{0}/J-\sigma/J)$ plane with constant critical temperature. In Fig. 4a, solid and dotted lines correspond to continuous and reentrant phase transition regions (of second and first order), respectively.  As we can see from Fig. 4a reentrant behavior can be observed only for specific set of $h_{0}/J$ and $\sigma/J$ parameters. Fig. 4b is a schematic representation of the phase transition characteristics of
\begin{figure}[h!]\label{fig3}
\subfigure[\hspace{0 cm}] {\includegraphics[width=7.25cm]{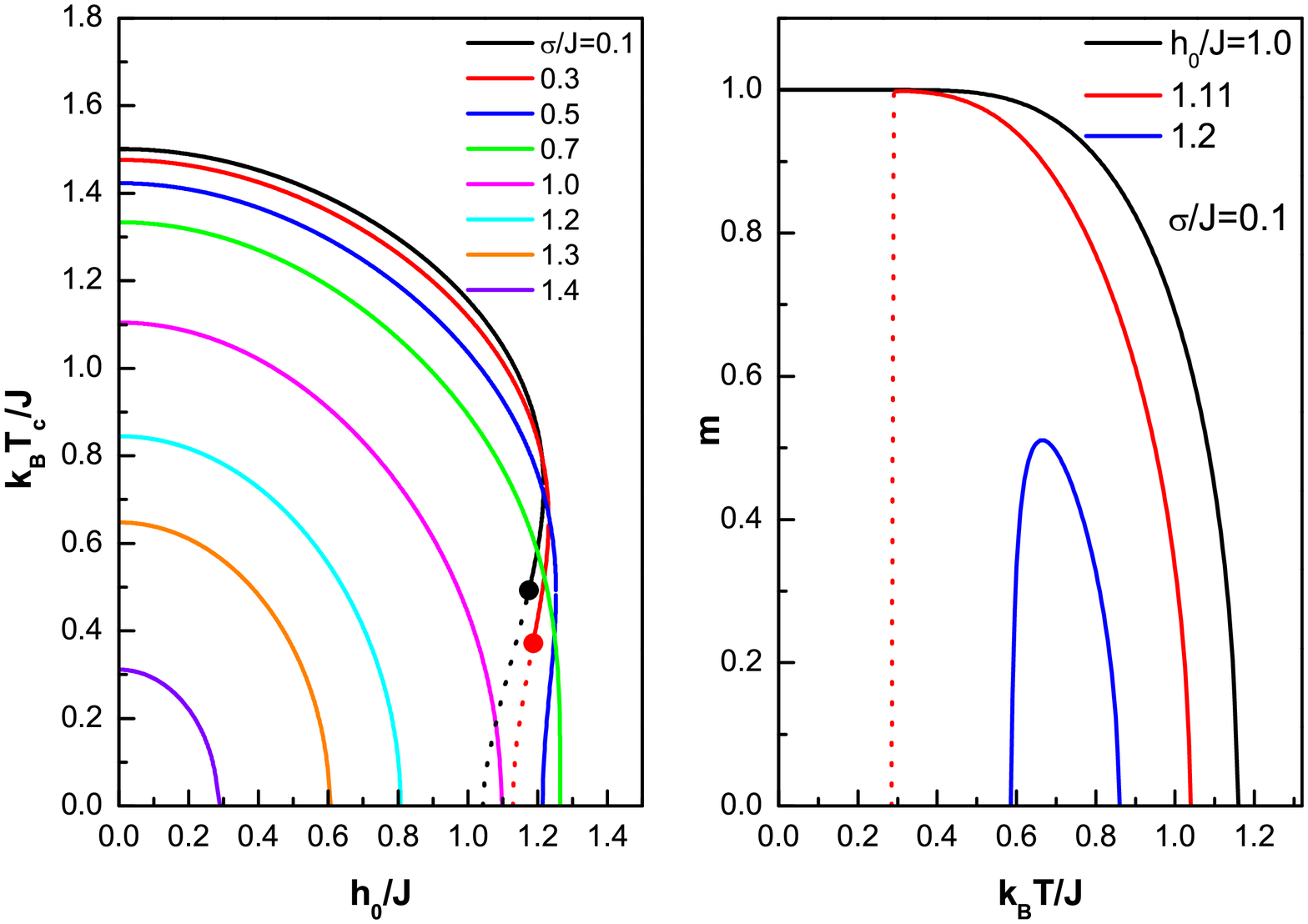}}
\subfigure[\hspace{0 cm}] {\includegraphics[width=7.25cm]{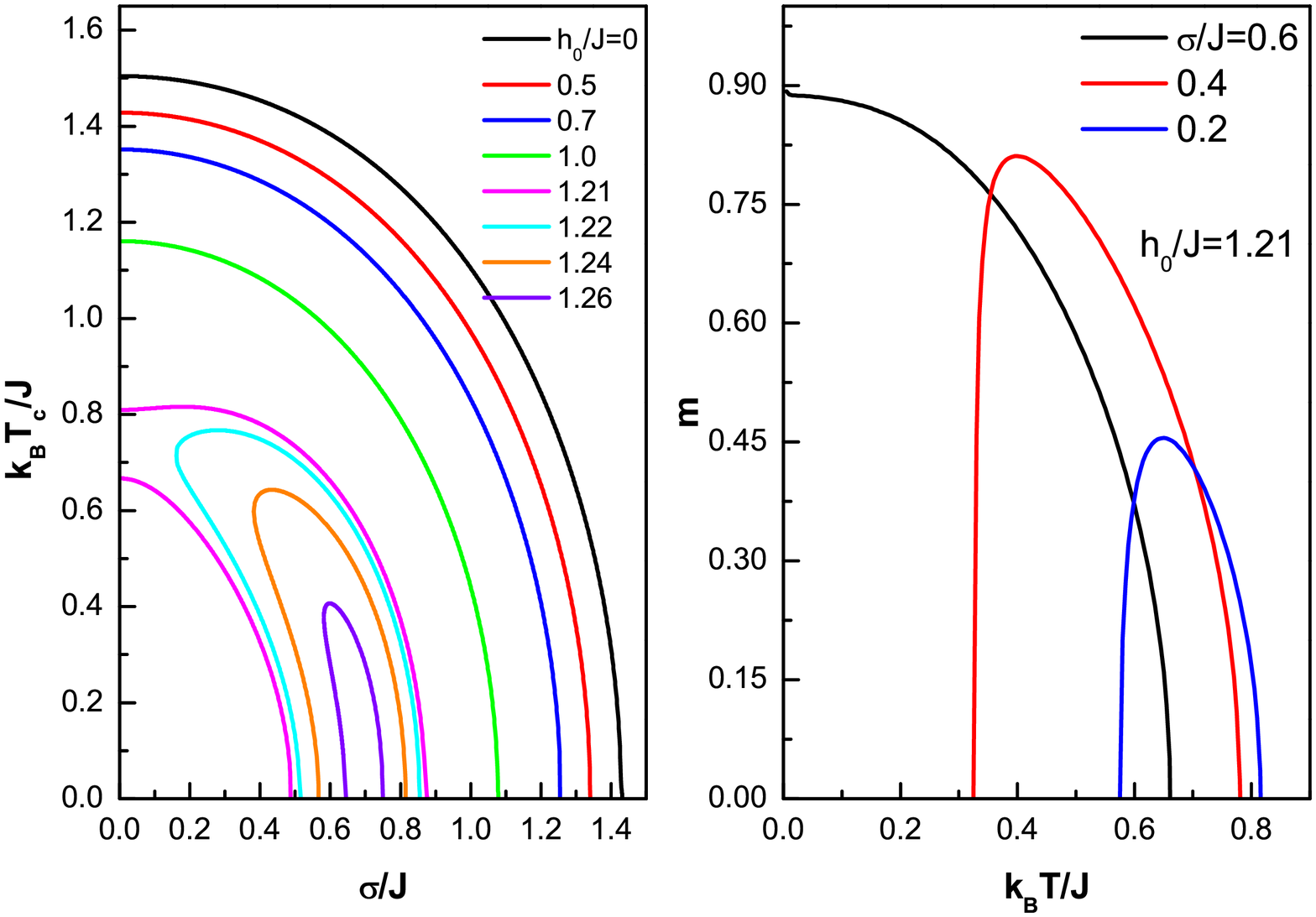}}\\
\center\subfigure[\hspace{0 cm}] {\includegraphics[width=7.25cm]{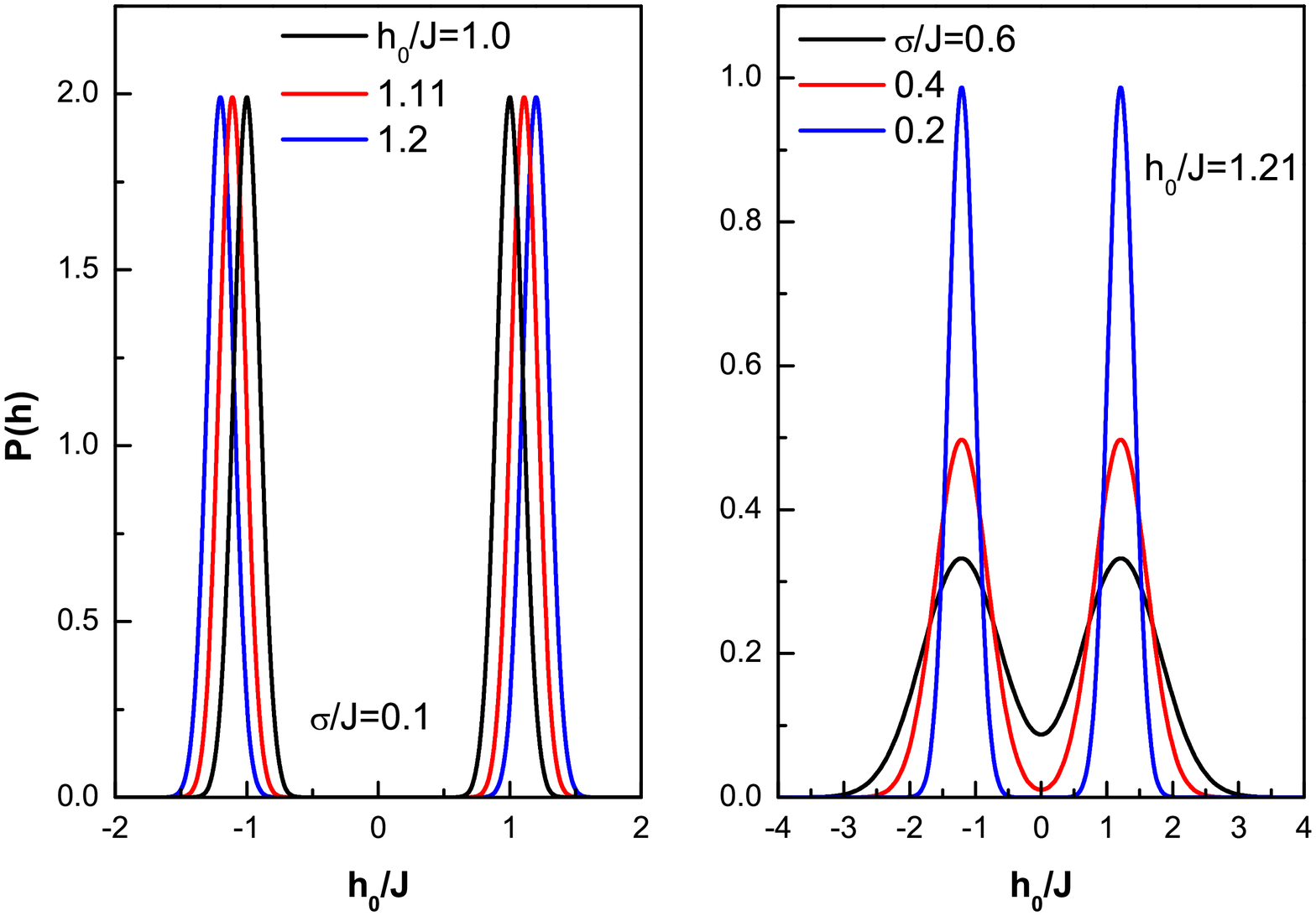}}
\caption{(a) Phase diagram (left panel) in $(k_{B}T_{c}/J-h_{0}/J)$ plane with selected values of $\sigma/J$ and variation of magnetization with temperature (right panel)  for selected values of $h_{0}/J$ with $\sigma/J=0.1$. Solid and dotted lines correspond to second and first order phase transitions and solid circles denote the tricritical points. (b) Phase diagram (left panel) in $(k_{B}T_{c}/J-\sigma/J)$ plane with selected values of $h_{0}/J$ and variation of magnetization with temperature (right panel)  with specific values of $\sigma/J$ for $h_{0}/J=1.21$. Solid and dotted lines correspond to second and first order phase transitions and solid circles denote the tricritical points. (c) Double Gaussian distribution corresponding to the magnetization curves depicted on right panels in  (b) and (c).}
\end{figure}
\newline
the system with double gaussian random field distribution. In this figure, the region which is colored in red represents $h_{0}/J$ and $\sigma/J$ values that cause a reentrant phase transition. If we select random field parameters $h_{0}/J$ and $\sigma/J$ from green colored region the system always undergoes only a secon order phase transition and similarly in blue region the system always remains in a paramagnetic phase for the whole temperature range. Furthermore, although increasing $\sigma/J$ value reduces the critical temperature of the system in general (see Fig. 4a), it is obvious in Fig. 4b that the green region where only second order phase transitions take place gets wider for the values of $\sigma<0.7$ and then it gets narrower for $\sigma/J>0.7$.

\section{Conclusions}\label{conclude}
In conclusion, we have studied the phase diagrams of a spin-1/2 Ising model in a random field by using three types of random field distributions. We have adopted an effective field approximation that takes into account the correlations between different spins in the cluster of considered lattice. As a consequence, when the random field distribution is single Gaussian the system always undergoes a second order phase transition between paramagnetic and ferromagnetic phases. This result agrees with those of the other works in the literature. For bimodal field distribution, the system exhibits reentrant phenomena. On the other hand, our numerical analysis clearly indicate that such a field distribution leads to a tricritical behavior and this observation contradicts with the results of previously published works based on EFT. In addition, in the absence of any randomness, numerical value of the critical temperature of the system obtained within the formalism of the present work agrees well with the exact result. Furthermore, double Gaussian form of the field distribution as well as single Gaussian counterpart has not been examined within the EFT approximation before. Hence we have also examined the system with this distribution function and we have obtained rich phase diagrams displaying reentrant phenomena and tricritical behavior. As a result, we can conclude that all of the points mentioned above show that our method is superior to the conventional EFT methods based on decoupling approximation and therefore the calculation results are more accurate.

We hope that the results obtained in this work may be beneficial from both theoretical and experimental points of view.
\begin{figure}[!h]\label{fig4}
\subfigure[\hspace{0 cm}] {\includegraphics[width=8cm]{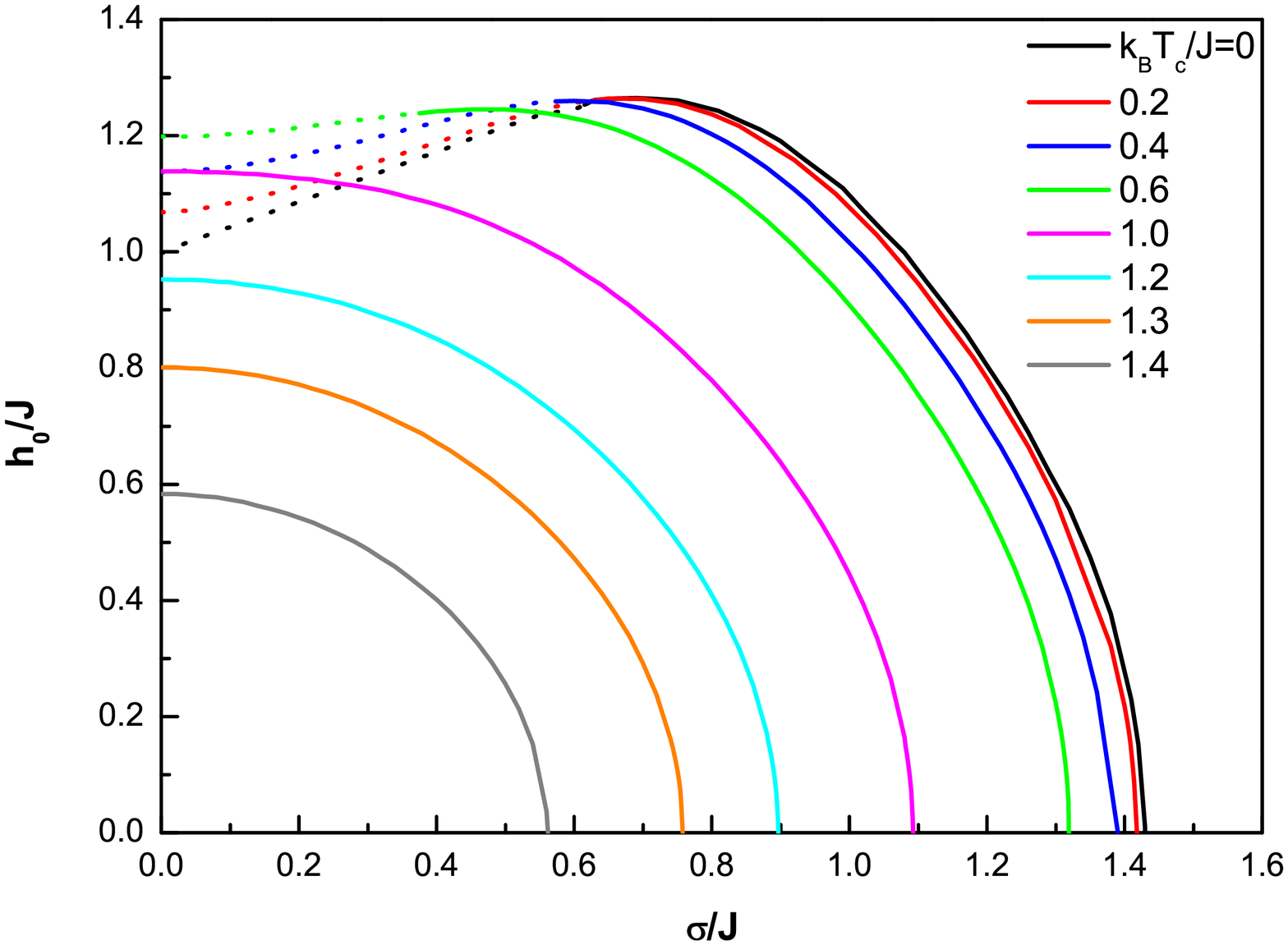}}
\subfigure[\hspace{0 cm}] {\includegraphics[width=7cm]{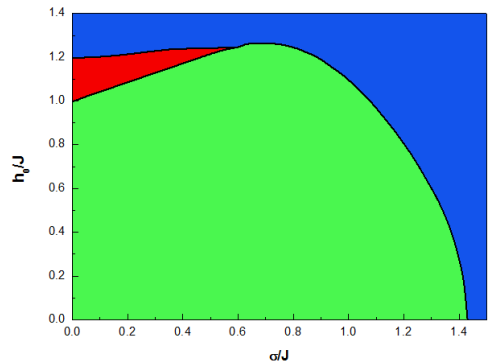}}\\
\caption{(a) Phase diagram in $(h_{0}/J-\sigma/J)$ plane with constant critical temperature, (b) a schematic representation of the phase transition characteristics of the system with double gaussian random field distribution. The regions with different colors indicates different phase transition characteristics.}
\end{figure}

\[\]
\[\]
\section*{Acknowledgements}
One of the authors (YY) would like to thank the Scientific and
Technological Research Council of Turkey (T\"{U}B\.{I}TAK) for
partial financial support. This work has been completed at the Dokuz
Eyl\"{u}l University, Graduate School of Natural and Applied
Sciences and is the subject of the forthcoming Ph.D. thesis of Y.
Y\"{u}ksel. The partial financial support from SRF
(Scientific Research Fund) of Dokuz Eyl\"{u}l University
(2009.KB.FEN.077) is also acknowledged.

\end{document}